%% file: main.tex
\documentclass[lettersize,journal]{IEEEtran} \def\TwoColumn{1}

\usepackage{xcolor}
\usepackage{amsmath,amssymb,amsfonts,amsthm}
\usepackage{algorithmic}
\usepackage{graphicx}
\usepackage{dsfont}
\usepackage{pstricks}
\usepackage{afterpage}
\usepackage{comment}
\usepackage{cite}
\usepackage{url}
\usepackage{etoolbox}
\usepackage{hyperref}
\hypersetup{hidelinks=true}
\usepackage{textcomp}
\def\BibTeX{{\rm B\kern-.05em{\sc i\kern-.025em b}\kern-.08em
    T\kern-.1667em\lower.7ex\hbox{E}\kern-.125emX}}
\allowdisplaybreaks[1]
\usepackage{tikz}
\usetikzlibrary{arrows}
\usetikzlibrary{shapes}

\ifdef{\TwoColumn}{}{ \usepackage{setspace}\doublespacing}

\makeatletter
\def\@endtheorem{$\blacksquare$\endtrivlist\@endpefalse } %
\makeatother

\newtheoremstyle{thmstyle}%
{}%
{}%
{}%
{}%
{\bfseries}%
{.}%
{ }%
{}%
\theoremstyle{thmstyle}

\usepackage[overload]{textcase}

\usepackage{textcomp}
\def\BibTeX{{\rm B\kern-.05em{\sc i\kern-.025em b}\kern-.08em
    T\kern-.1667em\lower.7ex\hbox{E}\kern-.125emX}}

\begin{document}
\title{SCAMPER - Synchrophasor Covert chAnnel for Malicious and Protective ERrands}
\author{P. Krishnamurthy, R. Karri, F. Khorrami
	\thanks{P. Krishnamurthy, R. Karri, and F. Khorrami are with the
		Dept. of ECE, NYU Tandon School of Engineering, Brooklyn, NY 11201, USA.
		(e-mails: \{prashanth.krishnamurthy, rkarri, khorrami\}@nyu.edu).}
	\thanks{This work was supported in part by DOE NETL (DE-CR0000017).}
}

\maketitle

\begin{abstract}
  We note that constituent fields (notably the fraction-of-seconds timestamp field)
 in the data payload structure of the synchrophasor communication protocol (IEEE C37.118 standard)
 are overprovisioned relative to real-world usage and needs, lending themselves to abuse for embedding of covert channels.
 We develop the SCAMPER (Synchrophasor Covert Channel for Malicious and Protective ERrands) framework to exploit these overprovisioned fields for covert communication and show that SCAMPER can be applied for both malicious (attack) and protective (defense) purposes.
Through modifications of the timestamp field, we demonstrate that SCAMPER enables an attacker to accomplish surreptitious communications between devices in the power system to trigger a variety of malicious actions.
These timestamp modifications can be performed without having any impact on the operation of the power system.
However, having recognized the potential for this covert channel, we show that SCAMPER can instead be applied for defensive security purposes as an integrated cryptographic data integrity mechanism that can facilitate detection of false data injection (FDI) attacks.
We perform experimental studies of the proposed methods on two Hardware-in-the-Loop (HIL) testbeds to demonstrate the effectiveness of the proposed SCAMPER framework for both malicious and protective purposes.
\end{abstract}


\bstctlcite{IEEEexample:BSTcontrol}
\section{Introduction}
\label{sec:introduction}

The security of modern power systems and other cyber-physical systems (CPS) has been increasingly of vital concern and importance in recent years~\cite{singer2015stuxnet,alert2016cyber,khorrami2016cybersecurity,liang2017review,faheem2018smart,bhamare2020cybersecurity,wlazlo2021mitm,nafees2023smart}. One of the driving factors in increasing the importance of strong security measures is the growing interconnectedness and digital transformation of electrical grid infrastructure. The integration of advanced sensors, real-time monitoring systems, and automated control mechanisms has enhanced grid reliability and efficiency and greatly facilitated the remote connectivity, programmability, and configurability of embedded devices in the power grid. However, this digital evolution has simultaneously expanded the attack surface available to adversaries, creating new vulnerabilities that threaten critical infrastructure security.
For example, Phasor Measurement Units (PMUs), Phasor Data Concentrators (PDCs), Remote Terminal Access Controllers (RTACs) and Remote Terminal Units (RTUs), intelligent electronic devices (IEDs), and other operational technology (OT) components now routinely communicate over network infrastructure that extend beyond earlier air-gapped environments and often provide bridges (with security mechanisms) to Information Technology (IT) networks. This increased connectivity, while enabling sophisticated grid management capabilities, has created pathways for adversaries to potentially compromise power grid safety, reliability, and performance through cyber attacks targeting communication protocols and device software.

Supply chain vulnerabilities represent another crucial potential threat to power system security that demands urgent attention. Adversaries may infiltrate both hardware and software supply chains\cite{karresand2003separating,rajendran2010classification,khorrami2016cybersecurity,krishnamurthy2023multimodal}, introducing malicious modifications that remain dormant until triggered remotely or in response to some CPS state/event. Supply chain compromises can be particularly insidious since they can bypass firewalls or other traditional perimeter security measures and infiltrate into sensitive system infrastructure.

In particular, adversaries can exploit supply chain vulnerabilities to modify firmware/software stacks on embedded devices to enable covert channels. From an information-theoretic perspective, covert channels can be viewed as essentially exploiting an ``excess'' of expressiveness in underlying data formats, communication protocols, or physical phenomena~\cite{yilmaz2020communication}. In the context of communication protocols, this excess of expressiveness manifests as leeway or ``play'' in constituent fields of protocol messages that can be manipulated without unaware receivers detecting any anomalies. Such covert channels can be surreptitiously used for data transfer/exfiltration as well as for signaling between compromised devices, enabling sophisticated attacks such as triggering malicious actions by RTACs or protective relays, or launching coordinated false data injection (FDI) attacks that can destabilize power system operations~\cite{alcaraz2019covert}.

Machine-in-the-middle (MITM) adversaries~\cite{wlazlo2021mitm,sen2021investigating} pose an additional critical threat in CPS including power systems. The infiltration of these adversaries into sensitive OT networks is becoming increasingly feasible due to expanding connectivity between IT and OT network segments. MITM adversaries can attempt to accomplish several malicious objectives including snooping and data exfiltration, communications modification in transit, injection of spurious communications (e.g., false data injection, malicious commands), and disruption of network operations (e.g., denial-of-service or DoS attacks).

The several security challenges outlined above are especially pressing since the majority of commonly used OT communication protocols were originally designed without comprehensive security considerations. Many protocols that are critical to power system operations, including those used for synchrophasor communications, SCADA systems, and RTAC/relay communications, such as DNP3, Modbus, C37.118, and OPC-UA were developed over several decades when the security landscape was very different from the current day (e.g., significantly more air-gapped network structures, much lesser concern about supply chain threats). For example, false data injection (FDI) attacks, while now recognized as a crucial threat to power system stability and performance~\cite{liang2017review,li2020detection_book}, are relatively straightforward for adversaries to execute once they establish an MITM presence within the power system network. The OT communication protocols themselves typically lack built-in security mechanisms such as cryptographic integrity verification or authentication. In particular, the IEEE C37.118 standard for synchrophasor data transfer is a vital example in this context of a highly critical OT communication protocol in modern power systems~\cite{ieee2024standard} for which security considerations have been largely overlooked.

In this paper, we highlight one specific and previously unexamined covert channel vulnerability in the synchrophasor C37.118 protocol. We name this methodology SCAMPER (Synchrophasor Covert chAnnel for Malicious and Protective ERrands), reflecting its dual nature that it can be applied for both attack and defensive security applications. We demonstrate that this covert channel can be instantiated through manipulations of payload-embedded timestamps, and that these manipulations can be accomplished in a lightweight manner that has no impact on power system operation. Fundamentally, the underlying source of this covert channel vulnerability lies in the overprovisioning of the fraction-of-seconds field in the data payload structure of the C37.118 protocol relative to real-world usage and needs. Since this field is embedded within the payload, covert channel mitigations that consider the TCP timestamp~\cite{zander2007survey,xing2020netwarden} cannot detect or prevent the covert channel implemented by SCAMPER. In addition, while we primarily focus on this fraction-of-seconds timestamp field in this paper, SCAMPER can also leverage the other fields in the payload to increase effective covert channel data rates, such as the phasors and analog values.

In addition to identifying and analyzing this new covert channel in C37.118 communications, we show that this same covert channel can be repurposed for defensive security purposes to help detect MITM adversaries that mount FDI attacks against synchrophasor communications. For defensive applications, the covert channel is used to embed cryptographic data integrity checks computed as hashes of payload data over time windows. The embedded data integrity check enables real-time detection and flagging of FDI attacks launched by MITM adversaries who modify synchrophasor measurement data in transit. SCAMPER thereby enables adding a valuable embedded defensive security mechanism into synchrophasor communications to guard against FDI attacks. In this context, SCAMPER highlights the fundamental dual nature of covert channels in cybersecurity: while they can be exploited by adversaries to surreptitiously convey information and perform malicious activities, the same underlying covert channel mechanism once recognized and understood can be applied constructively for defensive security purposes.

Our key contributions include:
\begin{enumerate}
\item identification of the existence of a potential covert channel in synchrophasor communications (IEEE C37.118 standard) through manipulations of payload-embedded timestamps
and development of the SCAMPER framework that enables both attack and defense applications through these subtle timestamp modifications
\item analysis and characterization of the identified covert channel including possible data rates and the computational overhead for implementing the covert channel
\item application of SCAMPER to enable an MITM FDI detector approach for defensive security purposes by leveraging the covert channel to embed cryptographic data integrity checks
\item experimental studies on HIL testbeds to demonstrate the practical feasibility of our approach in both attack and defense contexts.
\end{enumerate}

This paper is organized as follows. Section \ref{sec:relatedwork} reviews related work. Section \ref{sec:threatmodel} describes the threat models for SCAMPER including both its attack and defense applications. Section \ref{sec:covertchannel} discusses the exploitation of the covert channel by SCAMPER for attack purposes. Section \ref{sec:application_defense} repurposes the covert channel for defensive security and shows how the covert channel can be used as a drop-in embedded data integrity check that can facilitate detection of MITM FDI attacks. Section \ref{sec:experiments} presents experimental results for both the attack and defense applications of SCAMPER. Concluding remarks are summarized in Section \ref{sec:conclusion}.

\section{Related Work}
\label{sec:relatedwork}

\input{related_work}

\section{Threat Model}
\label{sec:threatmodel}

Since SCAMPER can be applied for both adversarial and defensive security purposes in the context of synchrophasor-based power system communications, we summarize below the threat models separately for these two applications.

\subsection{Covert Channel Attack Threat Model}
We consider a power system network where two or more devices (PMUs, PDCs, RTACs, etc.) are communicating using the IEEE C37.118 standard protocol. For example, this includes power network topologies such as where PMUs transmit their measurements to a PDC that aggregates the measurements or where a PMU sends its measurements to an RTAC that uses it for local control. We consider an adversary that has either compromised one or more of the devices that transmit C37.118 messages or has compromised the communication network itself (e.g., being able to modify messages through an MITM interpolation).

\noindent{\bf Adversary's goal:} The goal of the adversary is to modify the fraction-of-seconds timestamp fields embedded inside the data payloads of the time series of C37.118 messages to surreptitiously send information to a local or remote listener which has read access to the network communications. In particular, this implies that the intended listener(s) could include devices that are not explicitly communicating using C37.118.
Successful communication of this information might be used for various adversarial purposes such as triggering of distributed denial-of-service (DDoS) attacks, malicious RTAC or relay actions, coordinated false data injection, or other malicious activities that require synchronized action across multiple system components.

\noindent{\bf Stealthiness objective:} The adversary wishes to achieve this surreptitious communication without impacting the normal operation of the power system. Furthermore, the adversary wishes to remain stealthy by ensuring that all modified messages are compliant with the IEEE C37.118 standard, that data rates and inter-packet timings are unmodified, and that the modifications do not introduce any power system performance/stability degradations.

\subsection{Defensive Security Threat Model:}
When applied for defensive security purposes, SCAMPER seeks to facilitate detection of FDI attacks that are launched by an adversary who has compromised the communication network (e.g., by inserting an MITM attack that modifies the synchrophasor messages being transmitted through the network). For this purpose, SCAMPER repurposes the covert channel enabled by the subtle modifications of the fraction-of-seconds timestamp field in C37.118 messages to embed a cryptographic hash of the data payload contents of the messages over a time window.

\noindent{\bf Defender model:} In its defensive security role, SCAMPER can be deployed by the CPS operator or security administrator using either of two methods: (a) software updates of the C37.118 drivers on the devices that use this communication protocol to include support for the SCAMPER-enabled cryptographic data integrity checks; (b) insertion of small hardware modules at the physical network ports of these devices to insert the data integrity checks on the fly during communication.

\noindent{\bf Integration objectives:} To ease integration into existing power system networks, it is desired that the defensive security mechanism must operate transparently with existing power system applications, maintaining backward compatibility with legacy infrastructure while providing enhanced security for upgraded components.
The integrity verification process should introduce minimal computational overhead and latency to avoid impacting real-time data processing and control applications.
Additionally, to further ease integration into existing power system networks, it is desirable that data integrity checks can be performed by passive ``read-only'' listeners that can access the network traffic instead of requiring the C37.118 receiver devices to necessarily be the ones to perform the data integrity checks.

\noindent{\bf Detection goals:}
When applied for defensive security, SCAMPER's goal is to detect FDI attacks launched by MITM adversaries who attempt to modify synchrophasor measurement data in transit. This includes identification of both individual message tampering and coordinated attacks across multiple C37.118 data streams. For real-world applicability, it is crucial that the system should achieve high detection accuracy for MITM FDI attacks while maintaining low false positive rates and robustness against adaptive adversaries who may attempt to evade the integrity protection scheme or circumvent the data integrity check defenses through replay attacks.

\section{Payload-Embedded Covert Channel in C37.118}
\label{sec:covertchannel}

\begin{figure*}[!t]
    \centering
    \includegraphics[width=\textwidth]{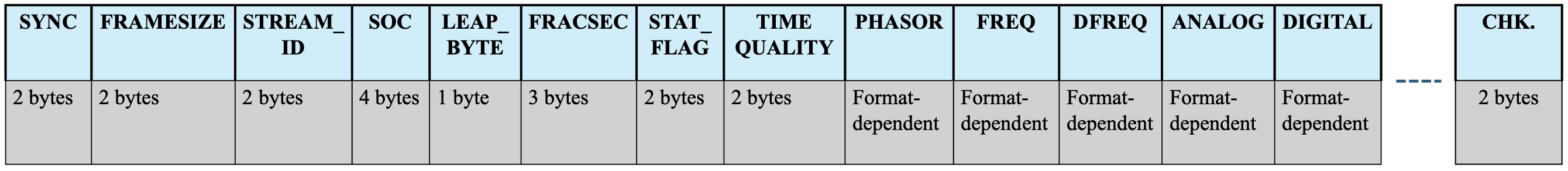}
    \caption{Data frame structure in the IEEE C37.118 standard.}
    \label{fig:c37_frame_structure}
\end{figure*}

In this section, we first provide an overview of the underlying mechanism of the covert channel and its application as an attack vector. In the next section, we will discuss the application of the covert channel mechanism for implementation of a defensive security methodology to detect MITM FDI attacks.

The data frame structure in the IEEE C37.118 standard~\cite{ieee2024standard} for periodic synchrophasor messages is summarized in Figure~\ref{fig:c37_frame_structure}. The fields in the data frame format and the corresponding numbers of bytes are shown in the figure. The meanings and roles of the fields are summarized briefly below:
\begin{enumerate}
\item SYNC: used for frame synchronization
\item FRAMESIZE: indicates the size of the transmitted frame in bytes
\item STREAM\_ID: identifier for the data stream
\item SOC: timestamp of the measurement as seconds since the Unix epoch (Jan 1, 1970)
\item LEAP\_BYTE: information about leap seconds
\item FRACSEC: fraction-of-seconds timestamp for the fractional part of the time (along with the seconds information in SOC)
\item STAT\_FLAG: status flags
\item TIME QUALITY: timestamp quality information
\item PHASOR: phasor measurement data (voltage or current)
\item FREQ: frequency measurement data
\item DFREQ: frequency rate-of-change measurement data
\item ANALOG: analog measurement data
\item DIGITAL: digital measurement data
\item CHK: cyclic redundancy check (CRC) checksum.
\end{enumerate}
Since measurements from multiple PMUs can be combined within a single C37.118 message, the fields from STAT\_FLAG to DIGITAL can be repeated as many times as the number of PMUs represented in the message.

For the covert channel addressed in this paper, the most relevant field is FRACSEC, which is the fraction-of-seconds timestamp measurement. The numerical value of this field is computed using the TIMEBASE setting which is specified in a configuration message and is 16777215 (0xFFFFFF) by default. Given a value of TIMEBASE, if the actual fractional part of the time is $t$, the value of FRACSEC is calculated as $\mbox{round}(t \times \mbox{TIMEBASE})$. To convert this back to the numerical fractional time, the corresponding calculation is $\mbox{FRACSEC}/\mbox{TIMEBASE}$.
The FRACSEC field is 3 bytes in length per the IEEE C37.118 standard. The overprovisioning of this field relative to real-world needs and usage is owing to the following reasons:
\begin{enumerate}
\item If TIMEBASE is the default value of 0xFFFFFF, the full range of values of FRACSEC from 0 to 0xFFFFFF will span the full range of valid values of the numerical fractional time (0 to $\sim$1). However, this means that the resolution of the fractional time is 1/0xFFFFFF, which corresponds to around 59.6 nanoseconds. However, the recommended accuracy of this timestamp is 1 microsecond~\cite{naspi2017time} and real-world accuracies can even be lower as seen, for example, in studies on time inaccuracies in synchrophasor measurements~\cite{shreshtha2023understanding,zadzar2023preventing}. Even assuming substantially higher accuracies in the future, it is reasonable to expect that at least the lowest bit of the FRACSEC field can be considered noise if the TIMEBASE is set to its default value.
\item On the other hand, a common setting of TIMEBASE in practice is 1000000, which corresponds to a resolution of 1 microsecond for the fractional time, which is the recommended timestamp resolution. However, this implies that the valid values for the FRACSEC field are limited to the range of 0 to 999999 since FRACSEC is the fractional part of the timestamp and is therefore less than 1 second. Since 999999 is 0xF4240, this implies that the highest 4 bits of the 3-byte FRACSEC field are unused. In fact, the IEEE C37.118 standard~\cite{ieee2024standard} provides example recommendations for TIMEBASE of values such as 720 since this is a multiple of the commonly used 60 Hz update rate for PMU measurements. Using such a lower value of TIMEBASE significantly constrains the possible valid values of FRACSEC implying that many more higher bits of the FRACSEC field are unused.
\end{enumerate}
Hence, in either case above, it is seen that the FRACSEC field being sized as 3 bytes is considerably overprovisioned relative to its actual real-world needs and usage. This overprovisioning opens the door for its use as a covert channel. Depending on the TIMEBASE setting being used, SCAMPER can utilize the lower or higher bits of FRACSEC (or a combination of both) as noted above to conduct its covert channel communications. In addition, while we focus on the FRACSEC field in this paper, SCAMPER can also utilize one or a few of the least significant bits of the phasor and voltage measurement values since these least significant bits are typically noise especially when a 4-byte format is used for these fields. This allows SCAMPER to further increase the effective covert channel bandwidth. With any combination of these bits being used by SCAMPER, it is to be noted that as part of the operation of SCAMPER, it would need to ensure that the checksum (the CHK field) in the C37.118 message is updated accordingly for consistency with the changes made to the FRACSEC field or other fields being used for the covert channel.

To calculate an estimate of the covert channel bandwidth, consider $k$ bits being used by SCAMPER per message (including in the FRACSEC and optionally phasor and analog values as noted above). With an update rate of $n$ messages per second for the PMU (and considering only one PMU in the C37.118 messages for simplicity), the covert channel bandwidth would be $kn$ bits per second\footnote{For simplicity, we simply consider the raw bandwidth and do not consider the overhead of a covert channel message frame format including its own checksum, etc. Standard message formats can be leveraged for this purpose with, for example, around 10-15\% overhead for a frame structure.}. For example, even with $k=1$ bit per message and $n=60$ messages per second, the covert channel bandwidth would be $60$ bits per second, which while low-bandwidth is more than sufficient for many covert channel objectives. In particular, since C37.118 messages are often sent to many devices such as RTACs (for local control), PDCs (for data aggregation), and HMIs (for visualization), and can also be inspected in transit by other devices (in promiscuous mode) on the same network segment, the covert channel can be used to deploy several types of malicious effects such as:
\begin{itemize}
\item Using the covert channel as a command and Control (C2) communications channel to send commands to RTACs/relays upon some power-related triggers to, for example, modify behavior, send malicious commands, launch DoS attacks, etc.
\item Performing surreptitious lateral movement, e.g., by using a PDC to hop between network segments
\item Performing data exfiltration of sensitive data such as passwords and operating parameters
\item Performing coordinated data injection attacks across multiple sensors
\item Using a PDC to distribute attack payloads/parameters to several devices
\item Propagate malware across several devices.
\end{itemize}

However, having recognized and characterized the covert channel, we believe that SCAMPER's more vital role is to repurpose this covert channel as an embedded data integrity check. This defensive security application of SCAMPER is discussed in the next section.

\section{Application to Cryptographic Protection against FDI}
\label{sec:application_defense}

The SCAMPER covert channel mechanism introduced in the previous section can be repurposed as an embedded mechanism for defensive security applications, particularly in protecting against FDI attacks mounted by MITM adversaries. By leveraging the overprovisioning of the FRACSEC field in the C37.118 protocol (and optionally additional least significant bits of phasor and analog signals as discussed above), we can establish a cryptographic integrity verification scheme without modifying the protocol specification.

\begin{figure*}[!t]
    \centering
    \includegraphics[width=0.7\textwidth]{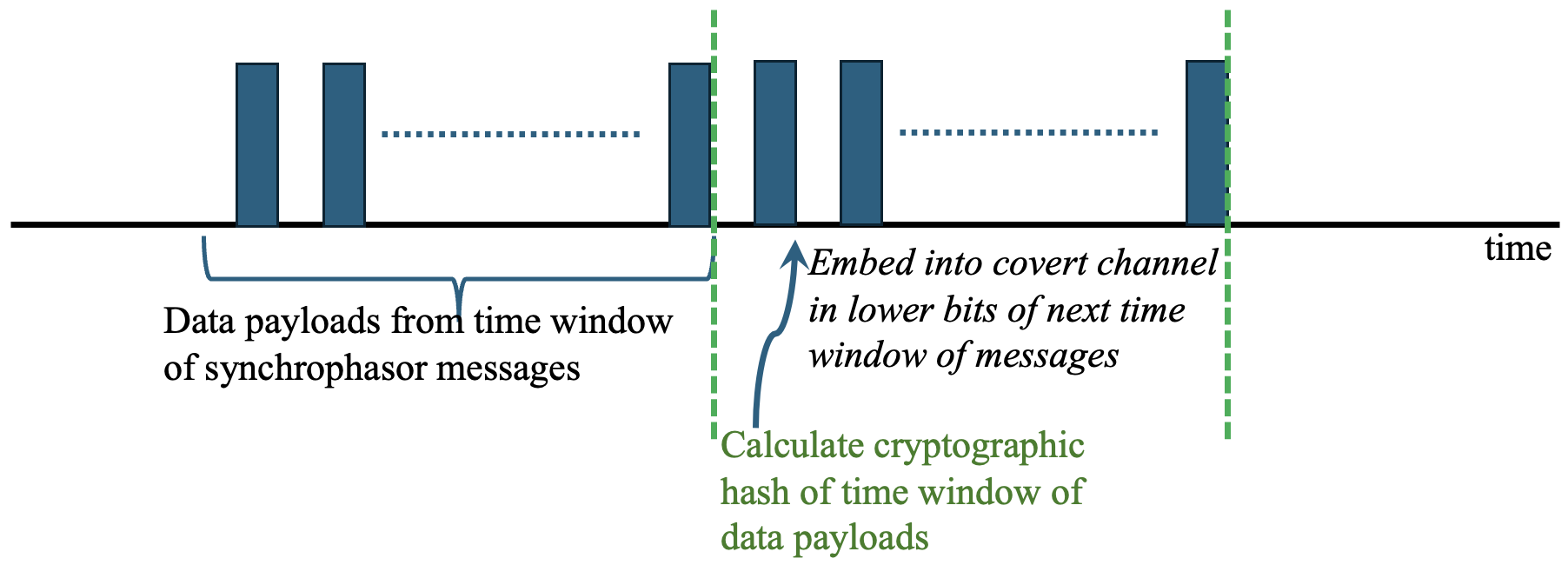}
    \caption{Application of SCAMPER for defensive security.}
    \label{fig:defensive_security_concept}
\end{figure*}

This embedded defensive security mechanism essentially leverages the fact demonstrated by the covert channel that there is an embedded additional information carrying capacity in the C37.118 communication protocol that can be used to embed a cryptographic hash of the actual message payloads for the purpose of data integrity verification. Since the available covert channel capacity is limited to only a few bits per message, we compute the hash over a time window of $N$ consecutive messages rather than attempting to embed a hash for each message. This hash is then distributed across the subsequent $N$ messages using the covert channel bits. This mechanism is illustrated in Figure~\ref{fig:defensive_security_concept}.

With the length of the cryptographic hash denoted as $h$ bits, if each message can carry $k$ bits of covert information, the time window is defined as $N = \lceil h / k \rceil$ since this is also the number of messages required to embed the hash. For example, consider a 128-bit hash (using, for example, Ascon as discussed further below). Also, consider the case where TIMEBASE is 1000000, which, as discussed in the previous section, gives us 4 unused bits in the FRACSEC field. In this case, we would require $\frac{128}{4}=32$ messages to transmit the complete hash. With a 60~Hz PMU update rate, this corresponds to a little more than half a second.

As illustrated in Figure~\ref{fig:defensive_security_concept}, the sequence of steps to embed this defensive security mechanism into a C37.118 message stream can be summarized as follows:
\begin{enumerate}
    \item During each time window of $N$ messages, insert the bits computed from the hash $H$ computed at the end of the previous time window (see Step 2 below). Also, during the time window of $N$ messages, store the data payloads (the complete messages or a subset of fields -- most importantly, the phasors, voltages, and currents) of the messages into a buffer $B$.
   \item At the end of the time window of $N$ messages, compute the hash $H$ of the contents of $B$ and a low-resolution timestamp and/or auto-incrementing counter (so as to prevent adversaries from mounting replay attacks); this hash $H$ is then transmitted using the covert channel over the next time window of $N$ messages.
\end{enumerate}

A suitable cryptographic hash for this purpose is Ascon, which is a NIST draft standard (NIST SP 800-232) providing a highly efficient and flexible algorithm for confidentiality, integrity, and authenticity. Ascon supports configurable length of hashes through its eXtendable Output Function (XOF) variant; for example,  Ascon-XOF128 provides 128-bit hashes. Ascon is computationally very lightweight and can be performed in at most a small fraction of a millisecond even on embedded devices with sparse computational capabilities. Since this hash needs to be computed only once every $N$ messages and the resulting hash value is used over the next time window of $N$ messages, the available amount of time for this computation is the update rate of the C37.118 data stream, which is typically 60 Hz or 120 Hz, which corresponds to an inter-message interval of approximately 16.67 ms or 8.33 ms, respectively. Since the time required for an Ascon hash computation is well within a millisecond, it is sufficiently lightweight for this purpose.

At the receiving end, the integrity verification process involves the following steps:
\begin{enumerate}
\item Over each time window of $N$ messages, store the received data payloads into a buffer $B$. Also, over the time window, extract the bits of the embedded hash (which is for the data payloads of the previous time window)
    \item At the end of the time window of received messages, compute the cryptographic hash $H'$ for the contents of $B$ along with the expected low-resolution timestamp and/or auto-incrementing counter used to prevent adversaries from mounting replay attacks. This hash $H'$ will be verified at the end of the next time window. Also, verify that the hash received over this time window matches the hash $H_{prev}'$ that was computed at the end of the previous time window.
\end{enumerate}

The proposed defensive security mechanism can be implemented using two approaches:
\begin{enumerate}
\item Embedding this defensive security mechanism into an upgraded software stack on the original C37.118 transmitter and receiver. This would be a more efficient approach over the long run, especially for newer power devices that are still under active support by their vendors.
\item Using additional hardware modules inserted at the physical network ports of the C37.118 transmitter and receiver devices. These lightweight modules essentially monitor for C37.118 messages and update them in transit to embed the defensive security mechanism. This approach can be deployed without the original C37.118 transmitter/receiver devices having to be aware of the data integrity check mechanism being applied.
\end{enumerate}

The computation of the data integrity checks to detect FDI attacks can be performed either at the original C37.118 receivers (e.g., the PDCs or RTACs) or can be performed by passive monitoring systems introduced into the power system OT network using a SPAN port for read-only network access such as, for example, in the TRAPS system in \cite{krishnamurthy2024tracking}. An advantage of integrating the FDI detection into a passive monitoring system is that this monitoring and anomaly detection can be performed alongside other existing anomaly detection techniques being applied and the resulting alerts can be integrated into their existing operator interface dashboards.

\section{Experimental Studies}
\label{sec:experiments}
\begin{figure*}[!h]
  \centering
  \includegraphics[width=0.32\textwidth]{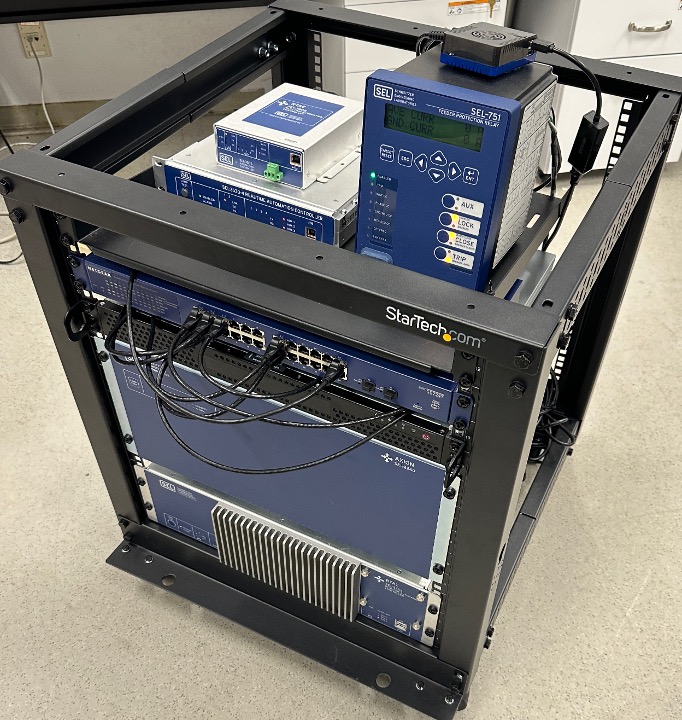}
  \includegraphics[width=0.55\textwidth]{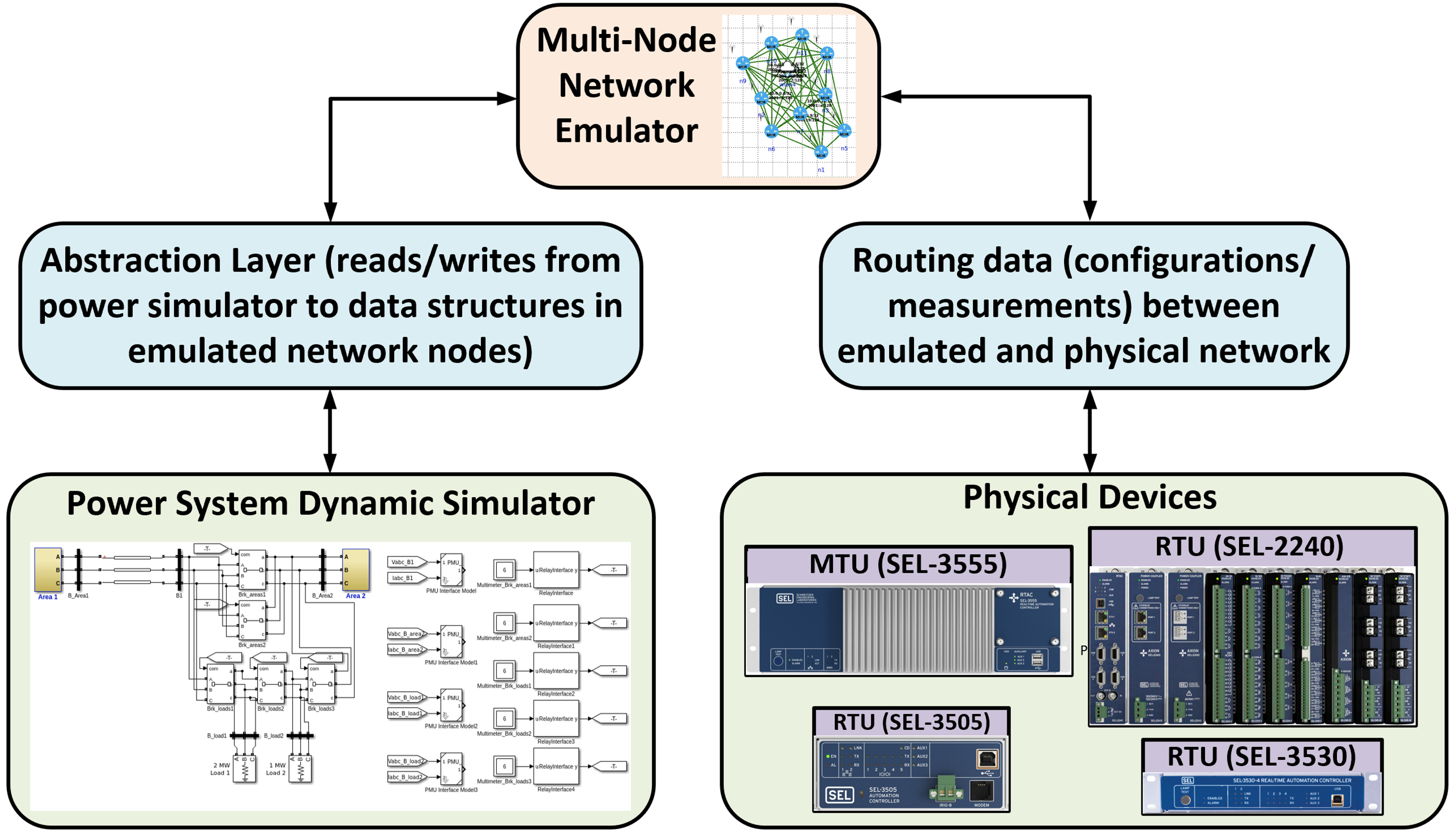}
  \caption{Smaller-scale HIL testbed (left) and its architecture (right).}
  \label{fig:nyu_testbed}
\end{figure*}

To study the effectiveness of SCAMPER, we perform experimental studies on two HIL testbeds. The first HIL testbed, which we will refer to as the smaller-scale HIL testbed in this paper, has the architecture shown in Figure~\ref{fig:nyu_testbed} and comprises of a set of physical devices
and virtual devices interfaced to a Matlab-based power system simulator and a
virtual network emulator. The physical devices in the testbed are four SEL (Schweitzer Engineering
Laboratories) RTACs (Real-Time Automation Controllers),
configured to have different roles including a HMI, data concentrator, and relay
control logic devices. Virtual devices are implemented via Linux namespaces
using the CORE (Common Open Research Emulator) network emulator tool for
building virtual networks. The virtual network emulator transparently routes traffic between both physical nodes and virtual nodes in the testbed.
The virtual devices are configured to simulate virtual IEDs such as Relays, PMUs, and PDC.
The power system dynamics simulator is implemented using Matlab and Simulink running on a Linux server.
All physical devices and the simulation computer are connected using an L2/L3 Netgear switch.
More details on this HIL testbed are in \cite{krishnamurthy2024tracking}.

\begin{figure*}[!htb]
  \begin{center}
    \includegraphics[width=0.95\textwidth]{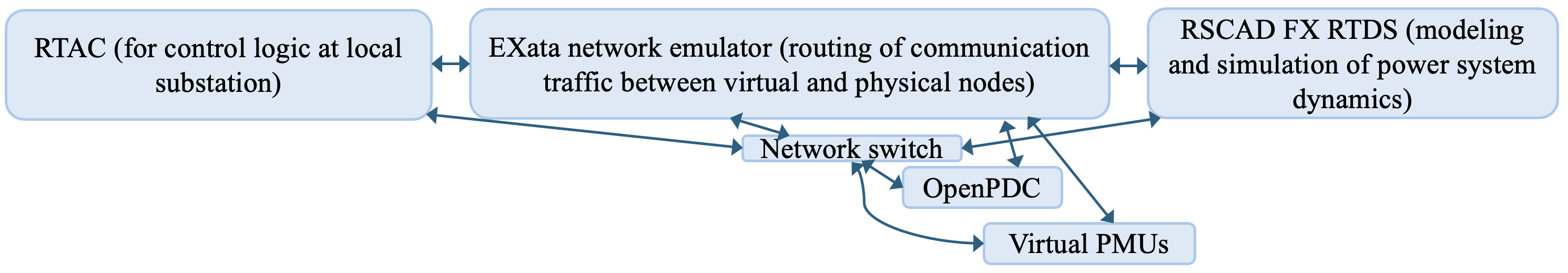}
  \end{center}
  \caption{Architecture of RTDS HIL testbed set up in NYPA's AGILe based on detailed modeling of a substation (with physical RTAC, connected PMU, virtual machine (VM) intruder, OpenPDC feed for other PMUs) in the context of the New York State power system topology.}
  \label{fig:nypa_testbed}
\end{figure*}

The second HIL testbed, which we will refer to as the RTDS HIL testbed in this paper, has the architecture shown in Figure~\ref{fig:nypa_testbed}, and is configured in the AGILe facility operated by NYPA (New York Power Authority).
This large-scale simulation setup includes a detailed model of a substation, with a physical RTAC, a connected PMU, and a virtual machine (VM) intruder interfaced via the EXata network emulator. This substation is simulated in the context of the complete New York State power grid model. The complete New York State power grid model simulated in the RTDS provides an additional 54 PMUs that are non-local to the substation modeled with greater detail. These additional PMUs represent signals from several locations in the New York State power grid dynamics model and are fed into an openPDC system that also provides a Historian interface. The primary subsystems of the simulation setup in AGILe are the RSCAD FX RTDS (Real Time Digital Simulator) for modeling and simulation of the power system dynamics, the physical RTAC for modeling of the control logic and configuration parameters as well as the HMI (Human Machine Interface), the OpenPDC data concentrator and Historian for monitoring the several PMUs from across the overall power system, and the EXata network emulator for routing of the communication traffic between the physical and virtual nodes. The network traffic in the testbed is captured using the EXata packet sniffing interface (PSI).

\begin{figure*}[ht]
\centering
\includegraphics[width=0.45\textwidth]{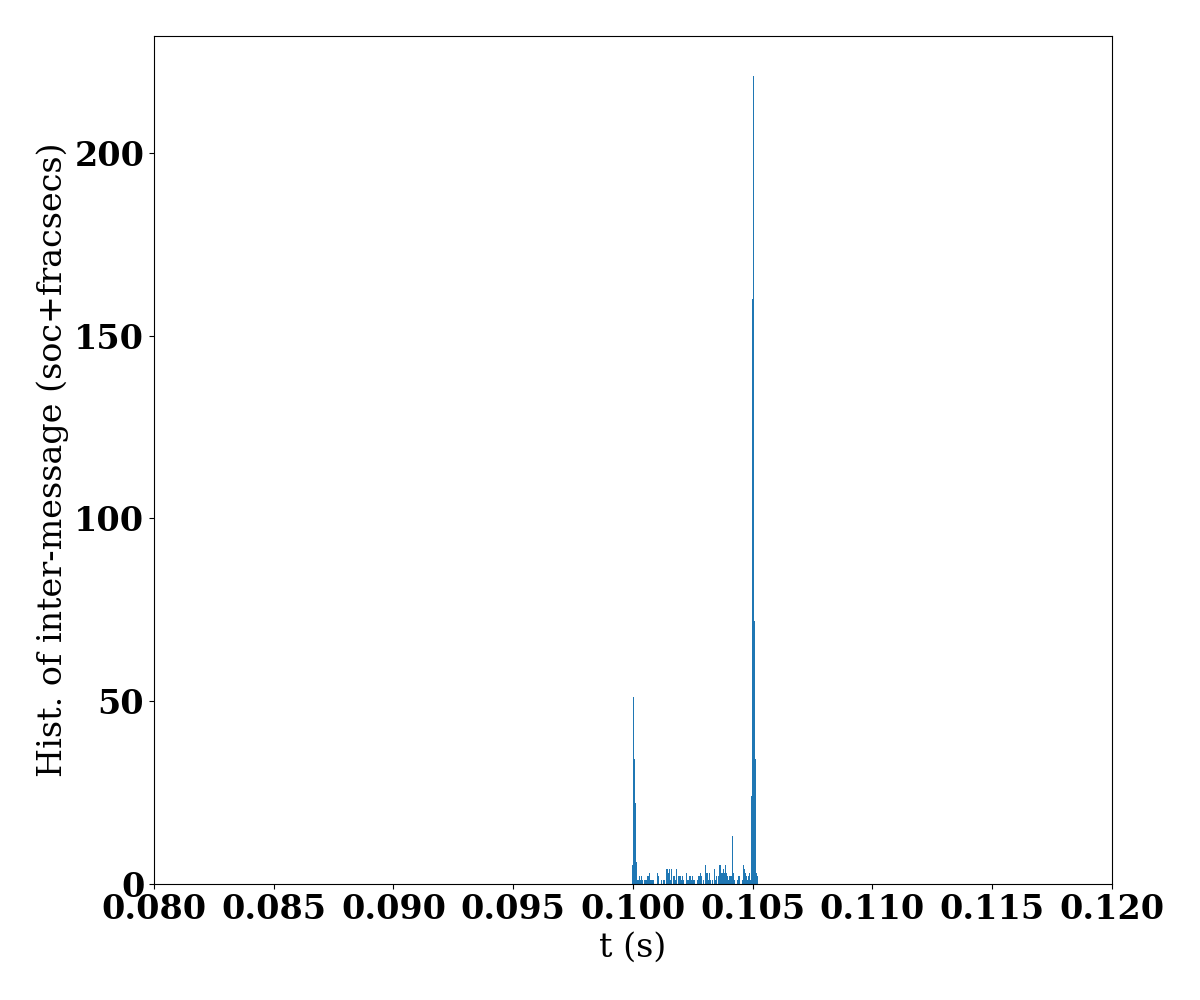}
\includegraphics[width=0.45\textwidth]{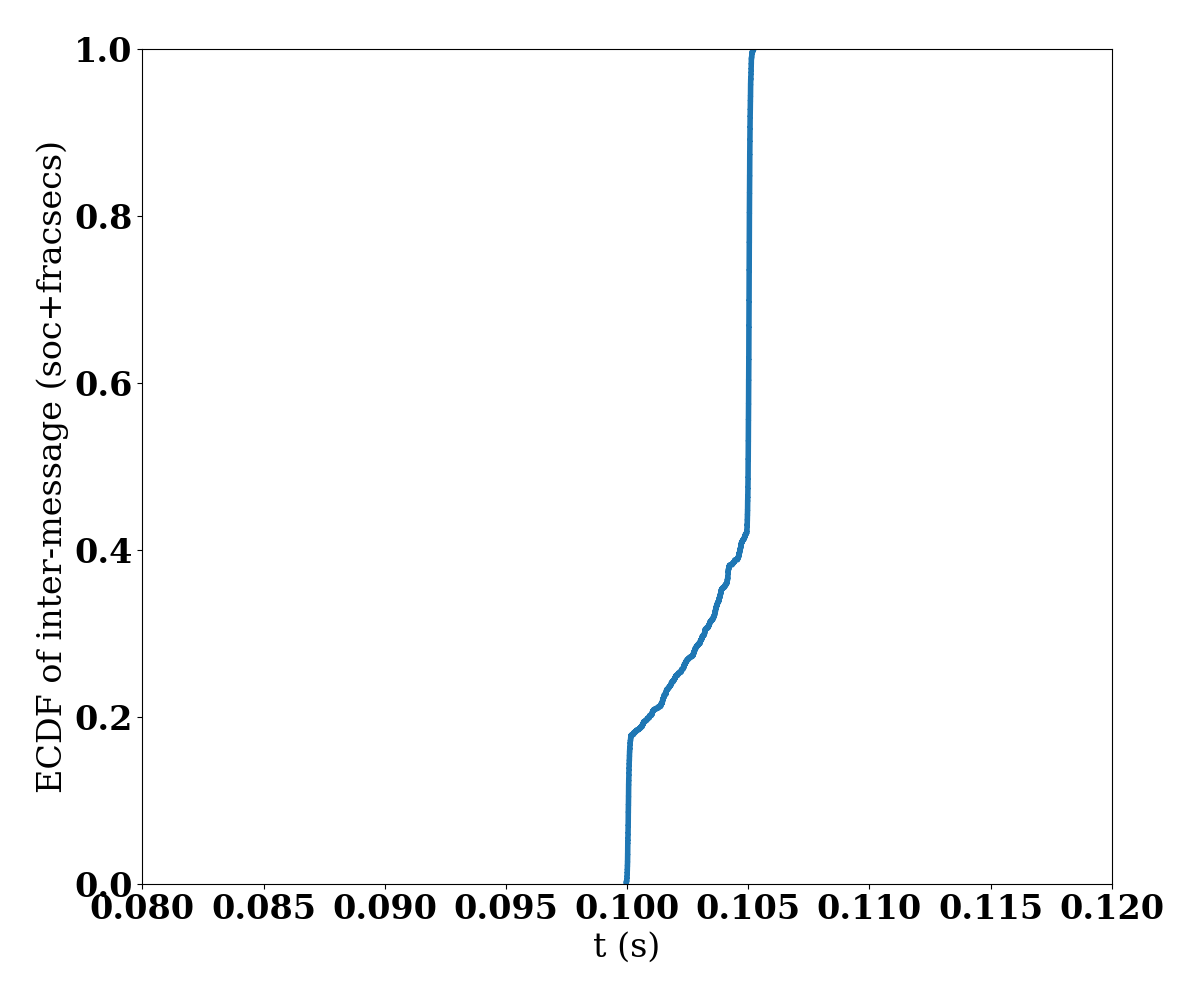}
\includegraphics[width=0.45\textwidth]{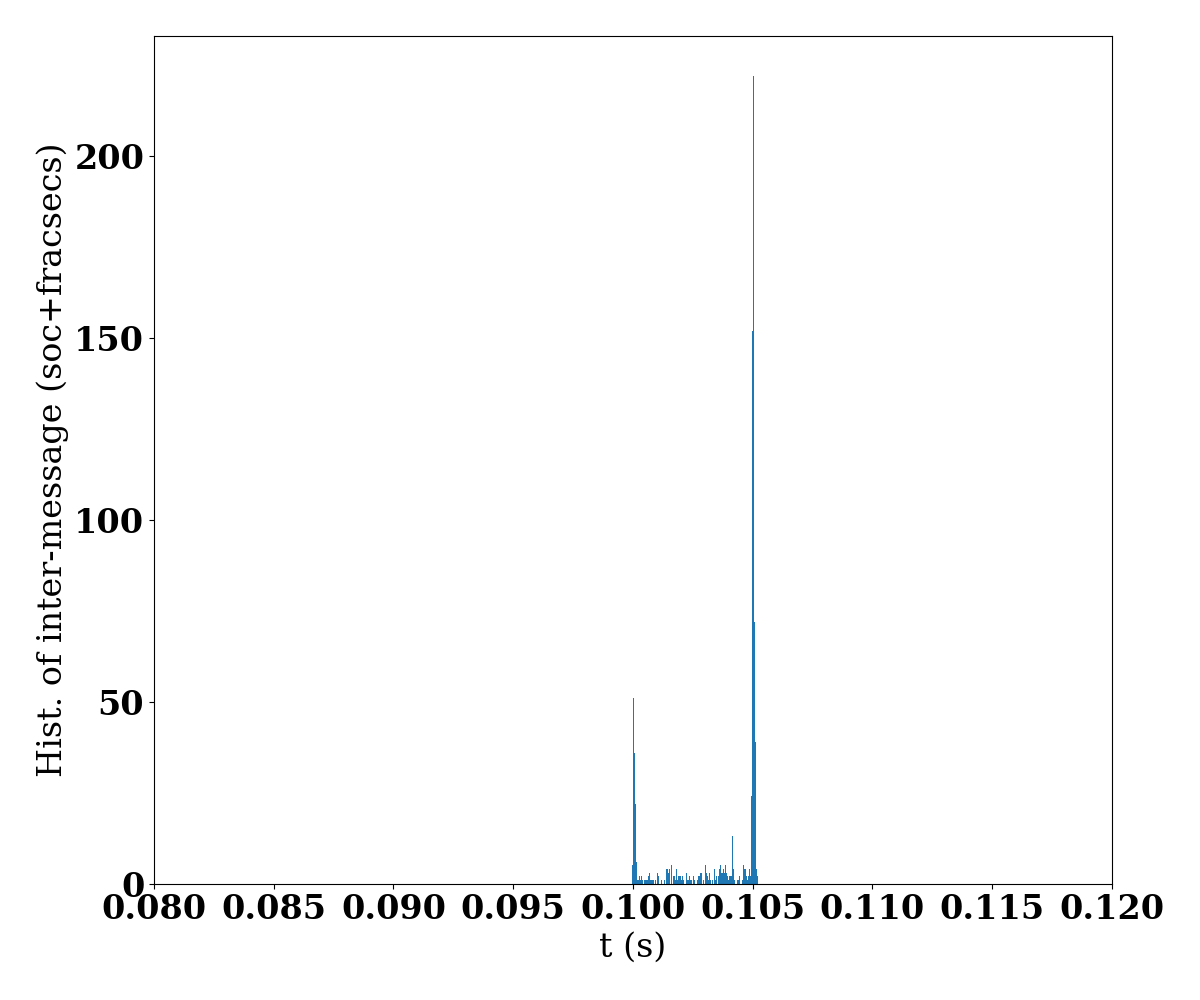}
\includegraphics[width=0.45\textwidth]{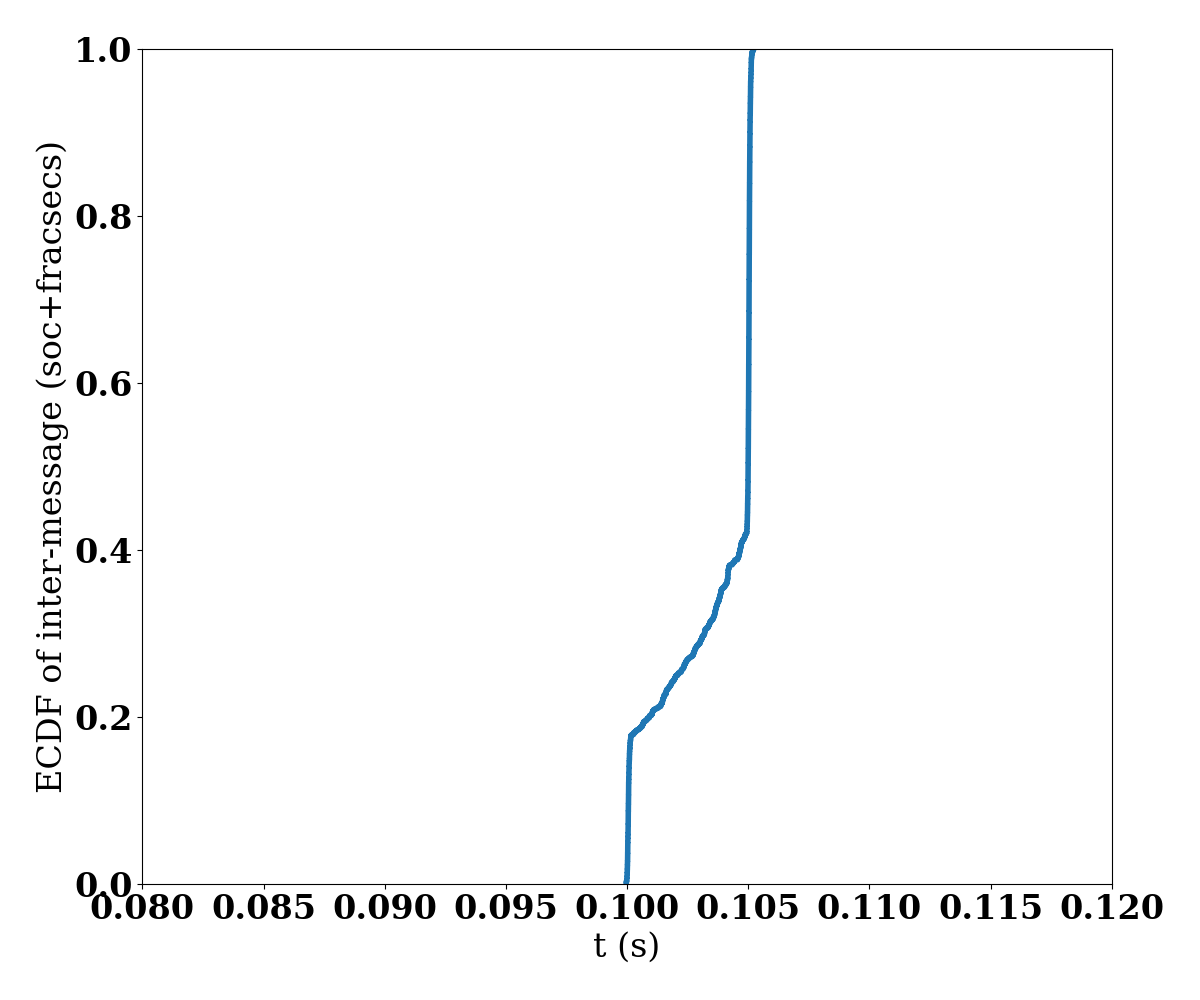}
\caption{Histogram (left) and eCDF (right) of the inter-sample times embedded in the message payload for a synchrophasor stream in the smaller-scale HIL testbed. The second row corresponds to the modified inter-sample times based on modifying the lowest bit of FRACSEC. The synchrophasor update rate is set to 10~Hz in this testbed.}
\label{fig:fracsecs_stats_hil}
\end{figure*}

\begin{figure*}[ht]
\centering
\includegraphics[width=0.45\textwidth]{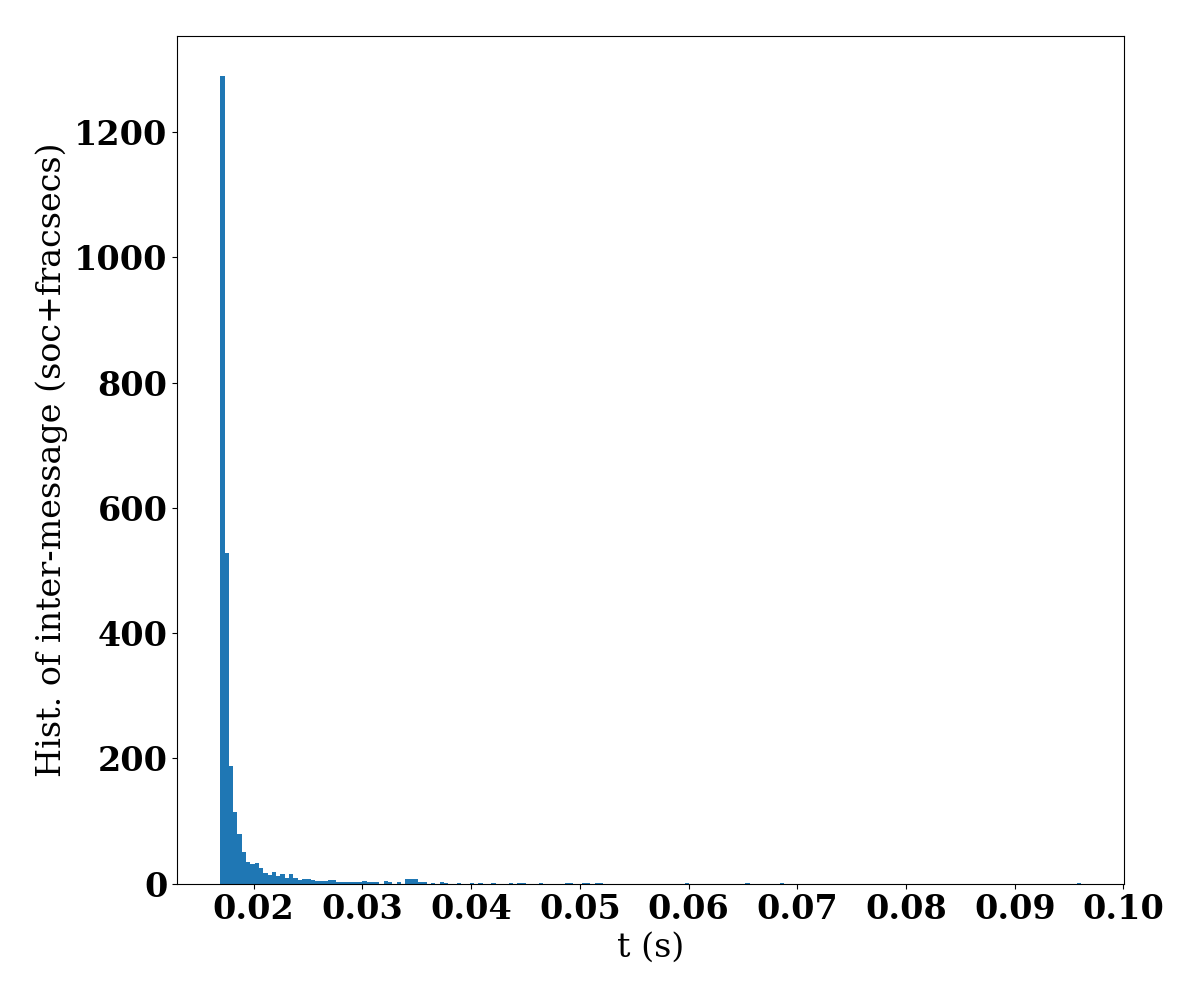}
\includegraphics[width=0.45\textwidth]{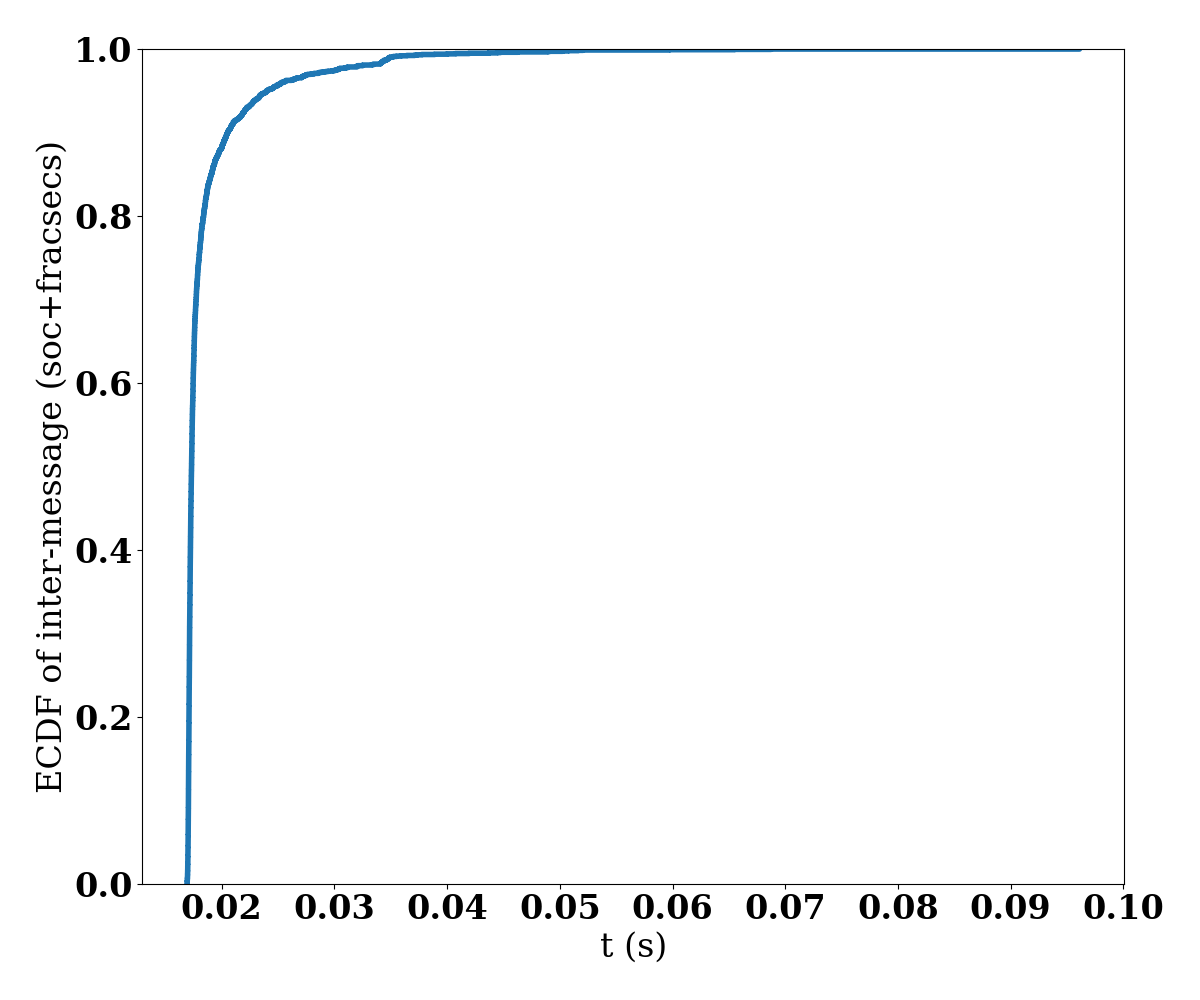}
\includegraphics[width=0.45\textwidth]{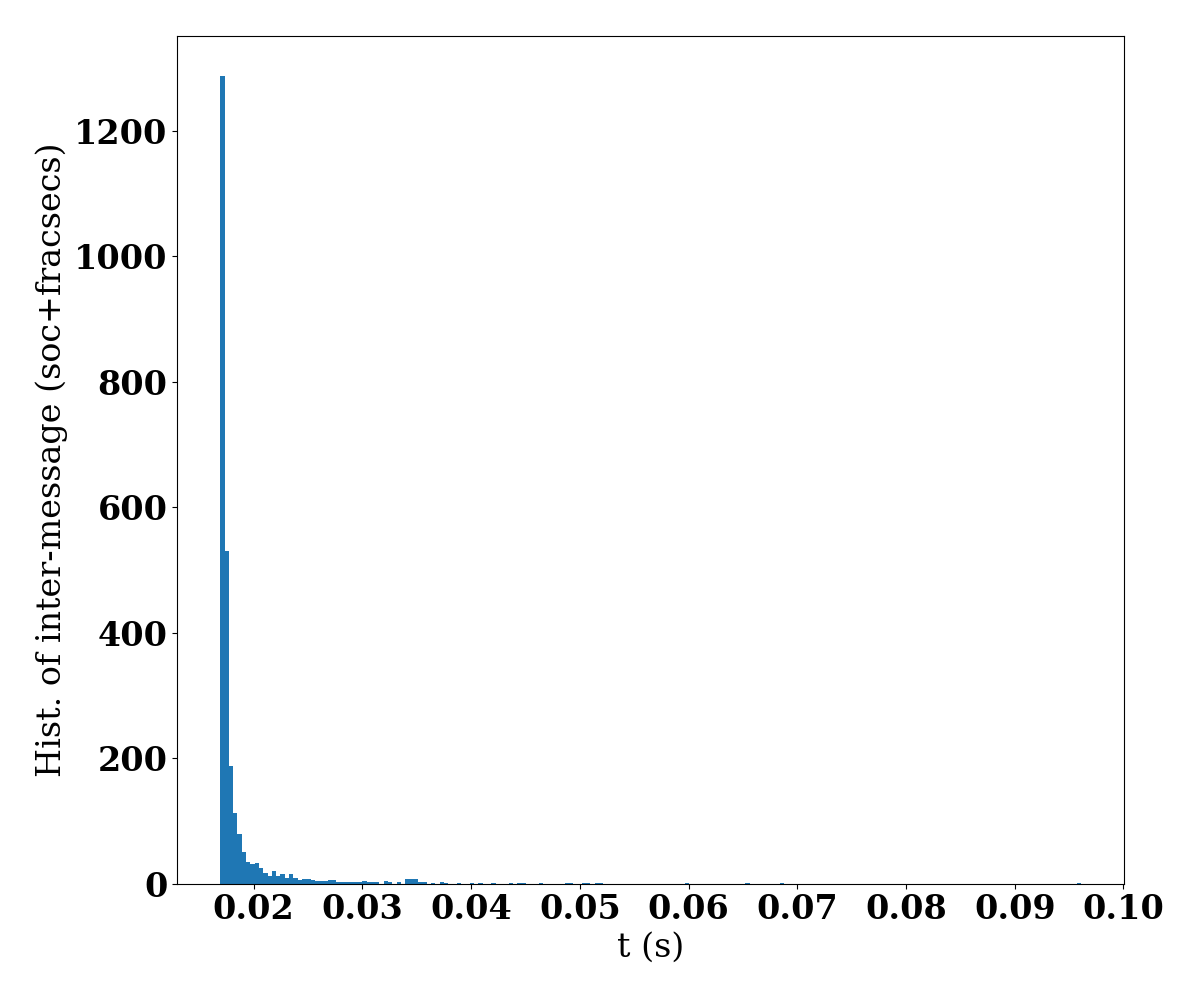}
\includegraphics[width=0.45\textwidth]{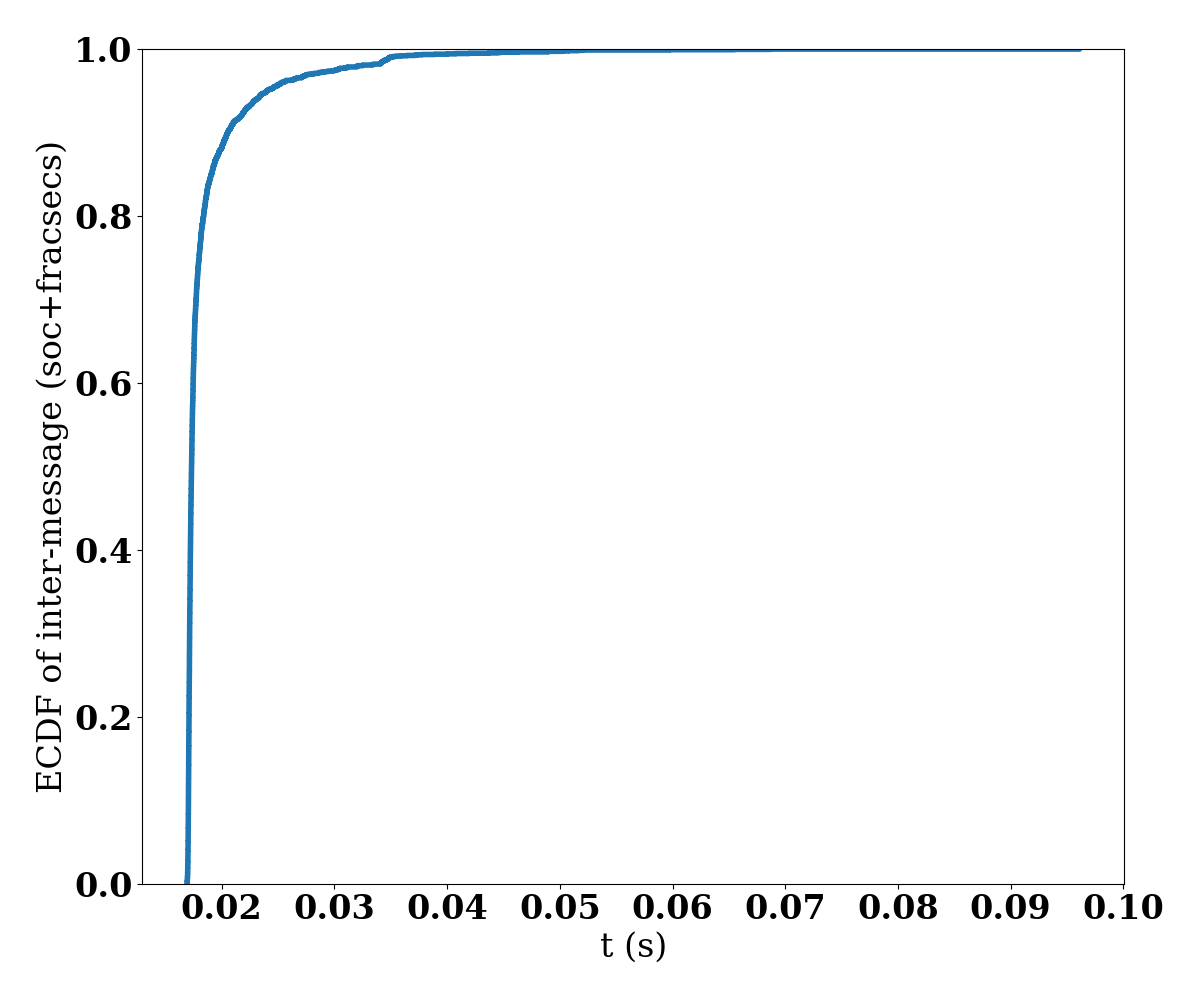}
\caption{Histogram (left) and eCDF (right) of the inter-sample times embedded in the message payload for a synchrophasor stream in the RTDS HIL testbed. The second row corresponds to the modified inter-sample times based on modifying the lowest bit of FRACSEC. The synchrophasor update rate is set to 60~Hz in this testbed.}
\label{fig:fracsecs_stats_rtds}
\end{figure*}

We will first consider the covert channel attack application of SCAMPER and then study the defensive security application. As discussed earlier, the specific bits that can be used to exploit the overprovisioning of the FRACSEC field in the C37.118 message format depends on the TIMEBASE setting. We consider here TIMEBASE set to 1000000, which is a commonly used setting since it corresponds to a 1 microsecond resolution of the effective fraction-of-seconds value, which is the recommended resolution. In this case, if the lowest (least significant) bit is used for the covert channel, one relevant question is whether modifications of the lowest bit would be within the typical noise tolerances of the inter-sample times. In Figures~\ref{fig:fracsecs_stats_hil}, we show the histogram and the empirical Cumulative Distribution Function (eCDF) of the inter-sample times embedded in the message payload (i.e., soc + fracsec where soc denotes the SOC seconds field and fracsec denotes FRACSEC/TIMEBASE) for a synchrophasor stream in the smaller-scale HIL testbed. The synchrophasor update rate is set to 10~Hz in this testbed. The analogous plots for a synchrophasor stream in the RTDS HIL testbed is shown in Figure~\ref{fig:fracsecs_stats_rtds}. The synchrophasor update rate is set to 60~Hz in this testbed. In both Figure~\ref{fig:fracsecs_stats_hil} and \ref{fig:fracsecs_stats_rtds}, the second row corresponds to the modified inter-sample times based on modifying the lowest bit of FRACSEC. While there could be slightly higher variability in the inter-sample times since these are from simulation testbeds on commodity hardware rather than from dedicated hardware, it is seen that the modifications are well within the typical variations of the original inter-sample times, making them undetectable. Also, note that when the lowest bit is used for the covert channel, the maximum change from its original value is 1 and often, the original value is kept unmodified since it could already happen to have the desired bit.

On the other hand, by using the unused higher bits of FRACSEC, SCAMPER can achieve much higher bandwidths with a larger message space as discussed earlier. In this case, the embedded modifications can be detected by a monitoring system that analyzes the FRACSEC field in the C37.118 messages. However, since this is a new covert channel that is not currently known, it is not detected by current-day monitoring systems. Nevertheless, as noted earlier, the more important application of SCAMPER is its ability to repurpose this FRACSEC field for defensive security applications. In this case, the undetectability of the modifications is not crucial. It is instead more important to ensure that the modifications do not interfere with the normal power system operations and that MITM FDI attacks are detected by SCAMPER's defensive security application. Both of these goals are achieved by SCAMPER's design since the FRACSEC modifications do not modify the normal data transfer using C37.118 and FDI modifications are indeed detected since the covert channel is used to communicate a cryptographic hash of time windows of the synchrophasor data payloads.

\begin{figure*}[!ht]
\centering
\includegraphics[width=\textwidth]{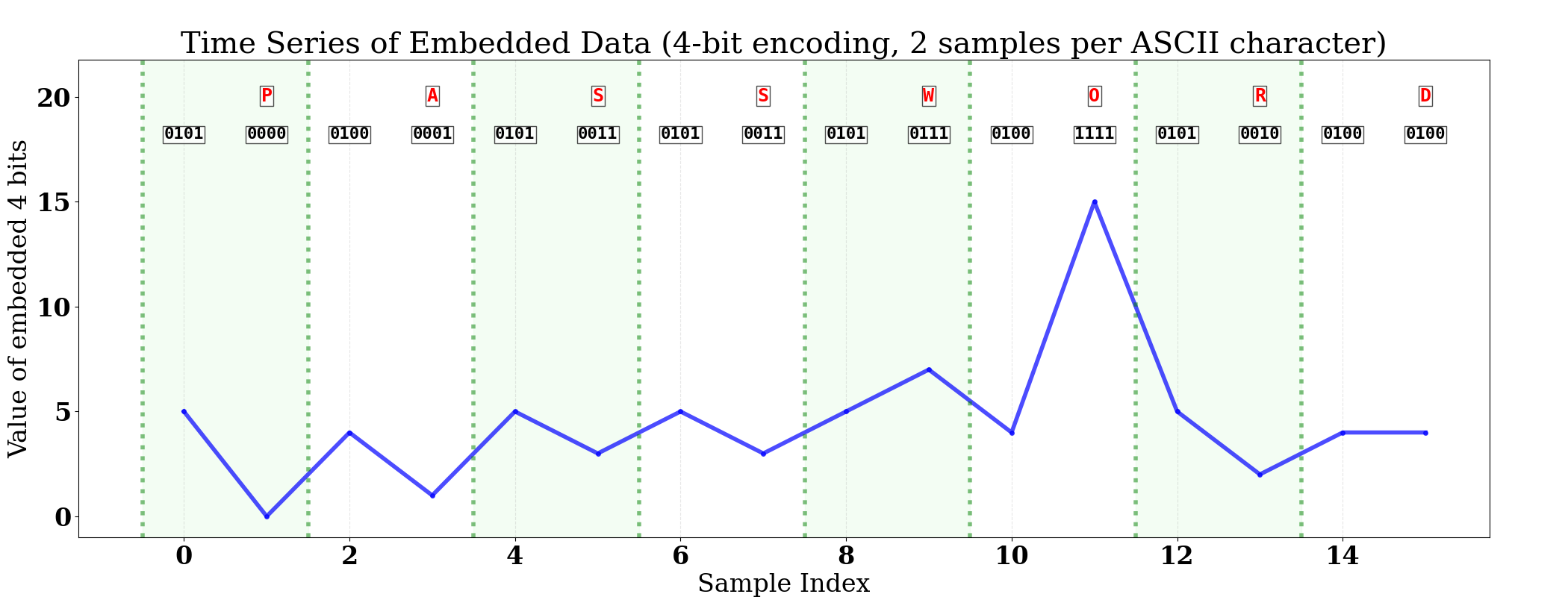}
\caption{Example of covert channel communication in the smaller-scale HIL testbed. The covert channel uses 4 bits per message and is configured to transmit the ASCII text "PASSWORD". The update rate is set to 10~Hz in this testbed (i.e., 10 sample index values correspond to 1 s).}
\label{fig:covert_channel_timeseries_hil}
\end{figure*}

\begin{figure*}[!ht]
\centering
\includegraphics[width=\textwidth]{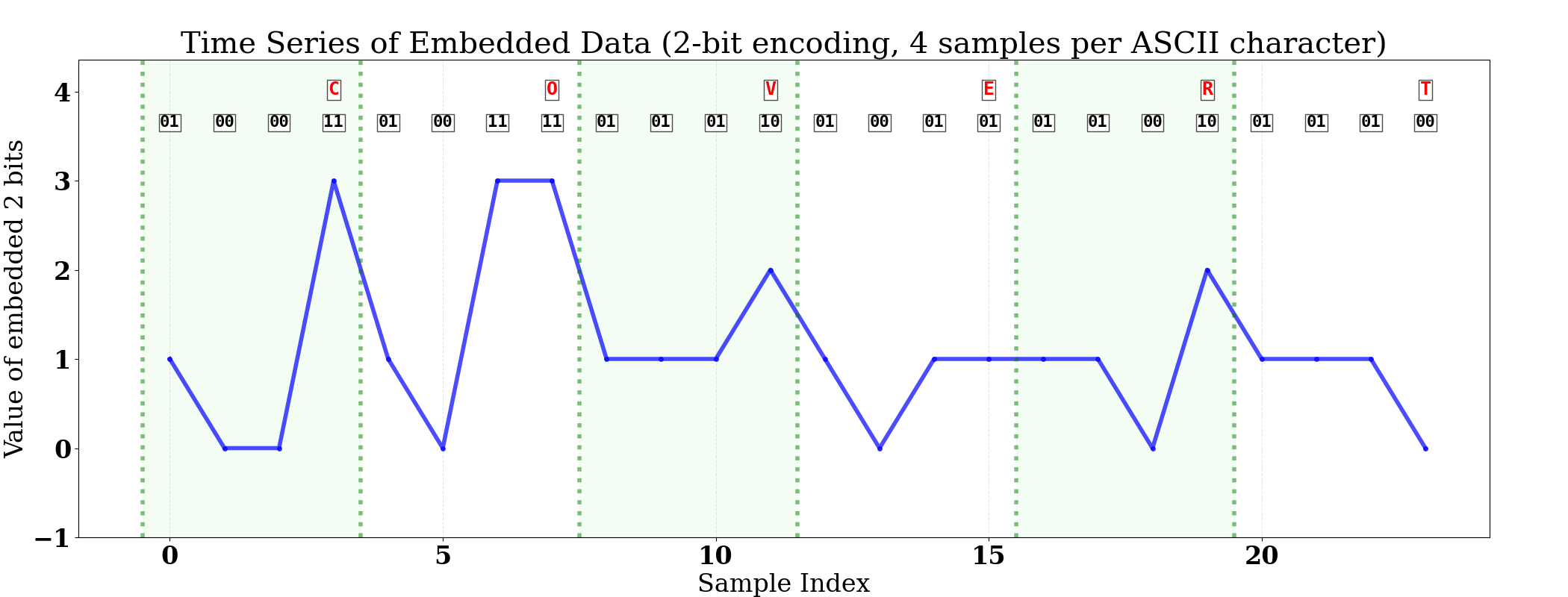}
\caption{Example of covert channel communication in the RTDS HIL testbed. The covert channel uses 2 bits per message and is configured to transmit the ASCII text "COVERT". The update rate is set to 60~Hz in this testbed (i.e., 60 sample index values correspond to 1 s).}
\label{fig:covert_channel_timeseries_rtds}
\end{figure*}

We first show examples of the covert channel communication in Figures~\ref{fig:covert_channel_timeseries_hil} and \ref{fig:covert_channel_timeseries_rtds} for the smaller-scale HIL testbed and the RTDS HIL testbed, respectively. The covert channel transmission is configured to use 4 bits per message in the SCAMPER deployment in the smaller-scale HIL testbed, while it uses 2 bits per message in the RTDS HIL testbed. For illustration, the messages being sent over the covert channel are specified as the ASCII text "PASSWORD" in the smaller-scale HIL testbed and "COVERT" in the RTDS HIL testbed.

\begin{figure*}[!ht]
\centering
\includegraphics[width=0.8\textwidth]{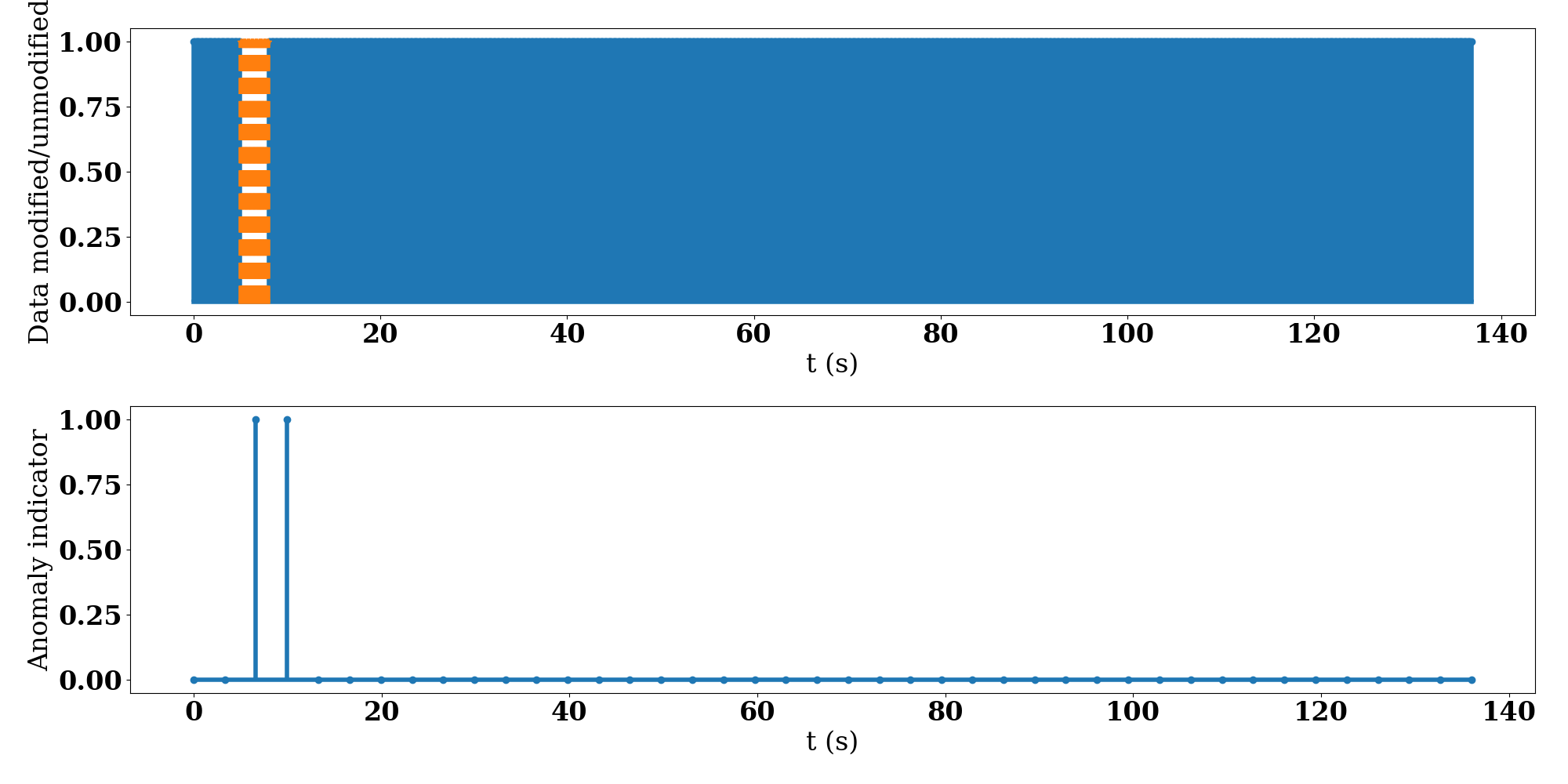}
\caption{Example of detection of an MITM FDI attack in the smaller-scale HIL testbed.  The first row shows the time interval of the injection of the MITM FDI attack (5 s to 8 s). The second row shows the anomaly detection time series based on validation of the cryptographic hash communicated over the covert channel over each time window (3.2 s with the settings for this study).}
\label{fig:fdi_detection_hil}
\end{figure*}

\begin{figure*}[!ht]
\centering
\includegraphics[width=0.8\textwidth]{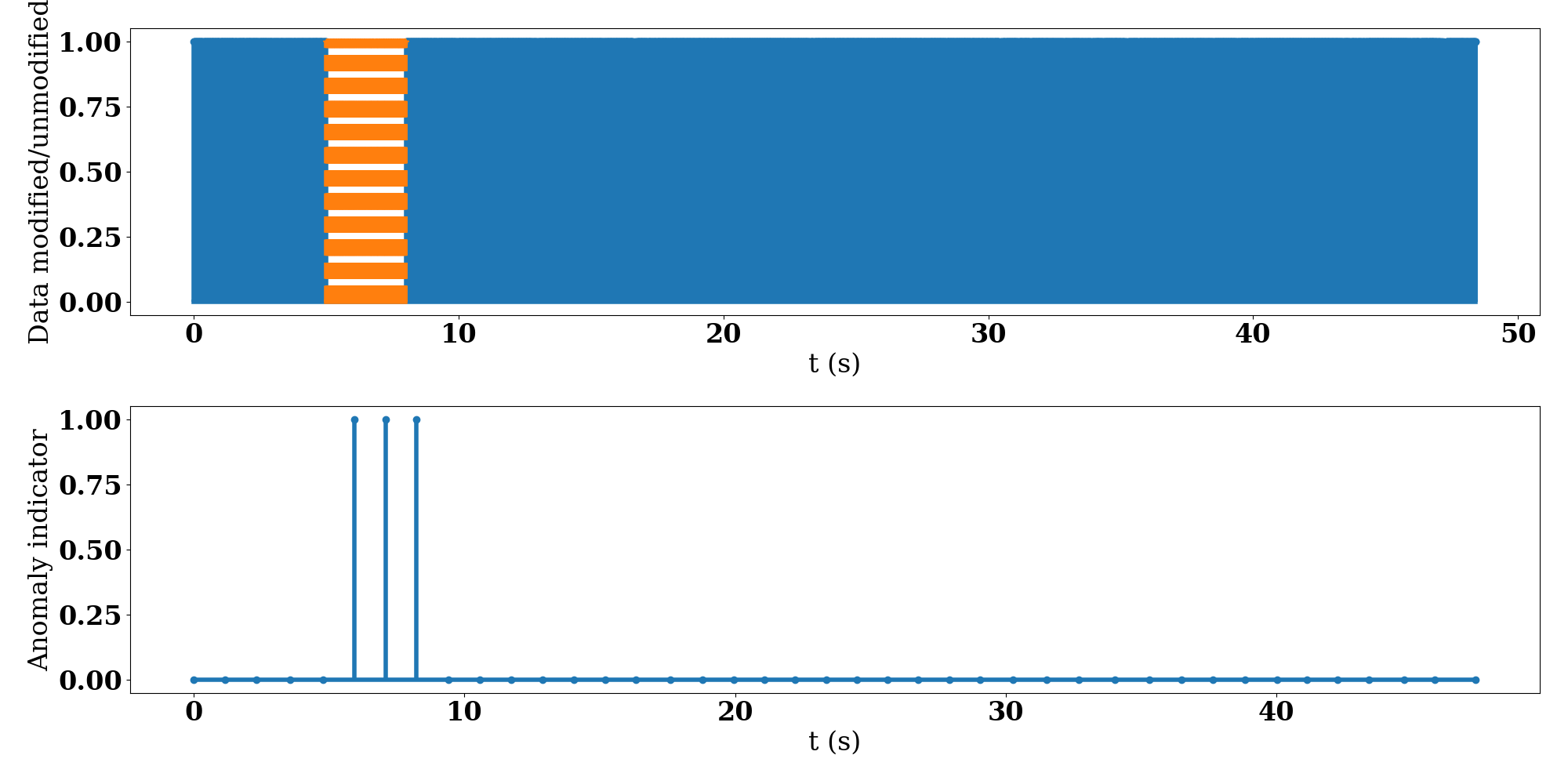}
\caption{Example of detection of an MITM FDI attack in the RTDS HIL testbed.  The first row shows the time interval of the injection of the MITM FDI attack (5 s to 8 s). The second row shows the anomaly detection time series based on validation of the cryptographic hash communicated over the covert channel over each time window (~1.1 s with the settings for this study).}
\label{fig:fdi_detection_rtds}
\end{figure*}

Next, to show the application of SCAMPER for defensive security, we show examples of detection of MITM FDI attacks in Figures~\ref{fig:fdi_detection_hil} and \ref{fig:fdi_detection_rtds} in the smaller-scale HIL testbed and RTDS HIL testbed, respectively. In both these cases, an FDI attack is injected from time 5 s to 8 s. A 128-bit Ascon hash is used for checking the data integrity. With 4 bits per message used for the covert channel in the smaller-scale HIL testbed, each time window is $\lceil 128 / 4 \rceil = 32$ messages, which with the update rate of 10 Hz in this testbed, corresponds to a data integrity check every 3.2 s. In the RTDS HIL testbed, with 2 bits per message used for the covert channel and an update rate of 60 Hz, each time window is $\lceil 128 / 2 \rceil = 64$ messages, corresponding to a data integrity check every around 1.1 s. These checks for a time window are performed based on the hash bits exfiltrated over the following time window as described in Section~\ref{sec:application_defense}. It is seen in Figures~\ref{fig:fdi_detection_hil} and \ref{fig:fdi_detection_rtds} that the FDI attacks over the time interval from 5 s to 8 s are flagged at each data integrity check for a time window that overlaps with any part of the FDI attack time interval. These anomaly detections are shown in the second rows of Figures~\ref{fig:fdi_detection_hil} and \ref{fig:fdi_detection_rtds}.

\section{Conclusion}
\label{sec:conclusion}

In this paper, we presented SCAMPER, a novel covert channel framework that exploits the overprovisioned fraction-of-seconds timestamp field in the IEEE C37.118 synchrophasor communication protocol. We showed that SCAMPER can be applied for both malicious attacks and defensive security and demonstrated its effectiveness using experimental studies on two HIL testbeds.
When applied for malicious purposes, SCAMPER enables attackers to establish covert communications between compromised devices in power systems through timestamp modifications that do not impact the power system operations.
More importantly, recognizing the possibility of this covert channel allows us to repurpose SCAMPER for defensive applications. We developed an integrated cryptographic data integrity mechanism that leverages the same timestamp field to detect FDI attacks launched by MITM adversaries.
More generally, our work highlights the critical importance of deeper examination of OT communication protocols from a security perspective, particularly identifying overprovisioned fields that may inadvertently create attack vectors. Our SCAMPER framework serves as both a cautionary example of potential vulnerabilities and a constructive approach to repurposing these vulnerabilities into defensive mechanisms. In this direction, future work will focus on more extensive analysis of SCAMPER's capabilities for both malicious and defensive applications, and on analysis of other OT communication protocols commonly used in industrial control systems, power grids, and critical infrastructure to evaluate applicability of similar covert channel techniques to these other OT protocols.

\vphantom{\cite{krishnamurthy2024tracking}}

\section*{Acknowledgments} \label{sec:Acknowledgments}
The authors would like to thank collaborators from Con Edison (Thai Thanh Nguyen) for their help in configuring the AGILe simulator testbed at NYPA and providing access and support for the testbed. The authors would also like to thank collaborators from Narf (Prashant Anantharaman, Michael Locasto, and others) and SRI (Nick Boorman, Ulf Lindqvist) for their work on parsers for OT network communication protocols.

\bibliographystyle{IEEEtran}
\bibliography{refs}

\end{document}

%% file: related_work.tex
Covert channels have been long recognized~\cite{zander2007survey,alcaraz2019covert} as a potential vulnerability across a range of computing and communication contexts including generic IT networks to cyber-physical systems. The basic underlying concept in covert channels involves establishing surreptitious communication pathways that exploit subtle modifications of legitimate communications or accompanying physical phenomena (side channels) to transmit information undetected by conventional monitoring mechanisms. A broad survey of early work on covert channels is provided in \cite{zander2007survey}, which considered classifications of covert channel techniques across different layers of the Open Systems Interconnection (OSI) model (e.g., link, network, transport) and different mechanisms (e.g., unused header bits, padding, checksum, time-to-live field, timing). The early work in this area primarily focused on TCP/IP-based communications and used mechanisms such as adding small delays on processing of TCP packets to use timestamps as covert channels~\cite{giffin2002covert} and manipulating header fields such as TCP initial sequence
number (ISN)~\cite{murdoch2005embedding}. IP covert channels using packet timing were studied in \cite{cabuk2004ip} including both TCP and UDP.
Covert channels have also been embedded at the physical layer using inter-packet gaps~\cite{lee2014phy}.

Covert channels have also been studied in the context of application-level protocols such as HTTP~\cite{brown2010covert}, Network Time Protocol (NTP) used for time synchronization\cite{hielscher2021systematic}, Internet Control
Message Protocol (ICMP) packet reply timing~\cite{lu2022timestamp}, and tunneling of a covert TCP session through a UDP traffic~\cite{oakley2020protocol}.
Routing protocols such as OSPF can also be exploited for covert channels using their protocol fields~\cite{schneider2023network}.
Increasing the robustness of TCP covert channels by including integrity checks and retransmission has been considered in \cite{bistarelli2024tcp}.
Also, amplification of effective covert channel data rates by methods such as compression using previously seen data has been studied~\cite{Wendzel2025dyst}.
Mitigations against covert channels include traffic normalization, capacity limiting, packet buffering, etc., as discussed, for example, in \cite{zander2007survey,alcaraz2019covert,xing2020netwarden}. Since such mitigations can incur network performance penalties, performance boosting methods using programmable data planes were considered in the NetWarden system proposed in \cite{xing2020netwarden} to preserve network performance while mitigating covert channels.
Additionally, the detection of classes of covert channels such as packet timing have been studied using machine learning (ML) models using features such as analysis of time intervals and payload lengths~\cite{han2020covert}, TCP payload entropy and network flow properties such as port numbers~\cite{li2024detecting}.

Apart from covert channels in network communications, physical phenomena (side channels) of various kinds have also been exploited for covert channels. These include, for example, voltage-based covert channels on FPGAs using the on-chip power distribution network~\cite{gnad2021voltage}, electromagnetic (EM) emanations resulting from software execution~\cite{yilmaz2020communication} or processor operating frequency variations~\cite{krishnamurthy2023multimodal}, acoustic emanations from peripherals such as fans~\cite{krishnamurthy2023multimodal} or electrical instrumentation in CPS~\cite{krishnamurthy2018processa, krishnamurthy2018processb,pearce2022detecting}, variations of communication bit rates~\cite{soderi2024connection}, and sensor and actuator signals~\cite{herzberg2019chatty}. Such side channels are of particular relevance for CPS since CPS inherently involve several kinds of physical instrumentation.
Stealthiness properties of covert channels in stochastic CPS has been studied from a dynamic system perspective in \cite{lucia2021covert}. Covert channels in specific OT communication protocols used in CPS have been addressed such as Modbus~\cite{kirdan2024modbus} and OPC UA~\cite{kirdan2024covert}.

Despite this vast body of research, communication protocols of synchrophasors used extensively in power systems have remained unexplored in the covert channel context. The IEEE C37.118 standard~\cite{ieee2024standard} for synchrophasor data transfer between power systems devices has become ubiquitous in modern electrical grid infrastructure. This protocol was developed and standardized over several years by an IEEE Working Group~\cite{martin2014overview}. Recent works have examined time synchronization vulnerabilities~\cite{zadzar2023preventing} and the impact of time inaccuracy on synchrophasor applications~\cite{shreshtha2023understanding}. However, the potential for exploiting overprovisioned fields within the synchrophasor data payload for covert communication has not been investigated.

The SCAMPER framework represents a novel contribution to the covert channel literature by: (1) identifying and characterizing the existence of a potential covert channel in synchrophasor communications through exploitation of the overprovisioned fraction-of-seconds timestamp field, (2) demonstrating a dual-purpose approach where the same covert channel mechanism can be applied for both malicious attacks (covert channels) and defensive security applications (embedded data integrity checks), and (3) presenting experimental studies on realistic HIL testbeds to demonstrate practical relevance and feasibility in power system environments. Unlike previous work that has focused primarily on traditional IT network protocols or generic industrial control systems, SCAMPER specifically targets the critical infrastructure domain of electrical power systems, addressing a significant gap in the cybersecurity analysis of the widely-deployed synchrophasor OT communication protocol.